\documentclass{Rinton-P10x7}
\newcommand{\app}[3]{Astropart.\ Phys.\ {\bf #1}, #3 (#2)}

\newcommand{\hepph}[1]{{\tt hep-ph/#1}}
\newcommand{\astroph}[1]{{\tt astro-ph/#1}}
\newcommand{\prep}[3]{Phys.\ Rep.\ {\bf #1}, #3 (#2)}
\newcommand{\plb}[3]{Phys.\ Lett.\ B\ {\bf #1}, #3 (#2)}
\newcommand{\npb}[3]{Nucl.\ Phys.\ B\ {\bf #1}, #3 (#2)}
\newcommand{\cpc}[3]{Comm.\ Phys.\ Comm.\ {\bf #1}, #3 (#2)}
\newcommand{\apj}[3]{Astrophys.\ J.\ {\bf #1}, #3 (#2)}

\newcommand{\prl}[3]{Phys.\ Rev.\ Lett. {\bf #1}, #3 (#2)}
\newcommand{\prd}[3]{Phys.\ Rev.\ D\ {\bf #1}, #3 (#2)}
\newcommand{\rmp}[3]{Rev.\ Mod.\ Phys.\ {\bf #1}, #3 (#2)}

\newcommand{\href}[2]{#1}

\newcommand{\dams}{\Delta a_\mu({\rm SUSY})}

\begin{document}

\title{Muon anomalous magnetic moment and\\
supersymmetric dark matter}

\author{\underline{E.~A.~Baltz}}

\address{Columbia Astrophysics Laboratory, Mail Code 5247, 550 W 120th
St., New York, NY 10027
\\E-mail: eabaltz@physics.columbia.edu
}

\author{P.~Gondolo}

\address{Department of Physics, Case Western Reserve
University, 10900 Euclid Ave.,\\ Cleveland, OH 44106-7079}


\maketitle

\abstracts{The anomalous magnetic moment of the muon has recently been measured
to be in conflict with the Standard Model prediction with an excess of
$2.6\sigma$.  Taking this result as a measurement of the supersymmetric
contribution, we find that at 95\% confidence level it imposes an upper bound
of about 500 GeV on the neutralino mass and forbids higgsino dark matter.  More
interestingly, it predicts an accessible lower bound on the direct detection
rate, and it strongly favors models detectable by neutrino telescopes.  Cosmic
ray antideuterons may also be an interesting probe of such models.}

\section{Experimental Measurement}

Recently, the Brookhaven AGS experiment 821 measured the anomalous magnetic
moment of the muon $a_\mu=(g-2)/2$ with three times higher accuracy than it was
previously known\cite{DATA}. Their result is larger than the Standard Model
prediction,
\begin{equation}
a_\mu({\rm exp})-a_\mu({\rm SM})=(43\pm16)\times10^{-10}.
\label{eq:disc}
\end{equation}
representing an excess of $2.6\sigma$ from the standard model value\cite{SM}.
One well-known possibility is that supersymmetric corrections to $a_\mu$ are
responsible for this discrepancy\cite{a_mu}.  We take the approach that all of
the measured discrepancy is due to supersymmetric contributions, and discuss
the implications for neutralino dark matter.  In this work we require\cite{prl}
\begin{equation}
10\times10^{-10}\le\dams\le75\times10^{-10}.
\end{equation}

\section{Supersymmetric Models}

The lightest stable supersymmetric particle in the Minimal Supersymmetric
Standard Model (MSSM) is most often the lightest of the neutralinos, which are
superpositions of the superpartners of the neutral gauge and Higgs bosons,
\begin{equation}
\tilde{\chi}^0_1 =
N_{11} \tilde{B} + N_{12} \tilde{W}^3 +
N_{13} \tilde{H}^0_1 + N_{14} \tilde{H}^0_2.
\end{equation}
We define the gaugino / higgsino ratio of the lightest neutralino as
\begin{equation}
\frac{Z_g}{1-Z_g} = \frac{ |N_{11}|^2 + |N_{12}|^2}{|N_{13}|^2 + |N_{14}|^2}.
\end{equation}

For large regions of the MSSM parameter space, the relic density $\Omega_\chi
h^2$ of the lightest neutralino is of the right order of magnitude for the
neutralino to constitute the dark matter in the Universe\cite{jkg96}. Here
$\Omega_{\chi}$ is the density in units of the critical density and $h$ is the
present Hubble constant in units of $100$ km s$^{-1}$ Mpc$^{-1}$. Present
observations favor $h = 0.7\pm 0.1$, and a total matter density $\Omega_{M} =
0.3 \pm 0.1$, of which baryons contribute roughly
$\Omega_bh^2\approx0.02$\cite{cosmparams}.  Thus we take the range
$0.052\le\Omega_\chi h^2\le0.236$ as the cosmologically interesting region, and
exclude models that do not satisfy this constraint.

We have explored a variation of the MSSM using the DarkSUSY
code\cite{DarkSUSY}.  Our framework has seven free parameters: the higgsino
mass parameter $\mu$, the gaugino mass parameter $M_{2}$, the ratio of the
Higgs vacuum expectation values $\tan \beta$, the mass of the $CP$--odd Higgs
boson $m_{A}$, the scalar mass parameter $m_{0}$ and the trilinear soft
SUSY--breaking parameters $A_{b}$ and $A_{t}$ for third generation squarks.
The only constraint from supergravity that we impose is gaugino mass
unification\cite{bg,coann,jephd}.  We have used a database of MSSM
models\cite{bg,coann,bub,neutrate,other_db} with one--loop corrections for the
neutralino and chargino masses\cite{neuloop}, and leading log two--loop
radiative corrections for the Higgs boson masses\cite{feynhiggs}. The database
contains a table of neutralino--nucleon cross sections and expected detection
rates for a variety of neutralino dark matter searches.  The database also
contains the relic density of neutralinos $\Omega_{\chi} h^2$, which includes
resonant annihilations, threshold effects, finite widths of unstable particles,
all two--body tree--level annihilation channels of neutralinos, and
coannihilation processes between all neutralinos and
charginos\cite{coann,GondoloGelmini}.

We examined each model in the database to see if it is excluded by the most
recent accelerator constraints. The most important of these are the LEP
bounds\cite{pdg2000} on the lightest chargino mass
and on the lightest Higgs boson mass $m_{h}$,
and the constraints from $b \to s \gamma$\cite{cleo}.

\begin{table}
\caption{The ranges of parameter values used in the MSSM scans of
Refs.~\protect\cite{bg,coann,bub,neutrate,other_db}.  We use approximately
79,000 models that were not excluded by accelerator constraints before the
recent $a_\mu$ measurement. }
\begin{center}
\begin{tabular}{|r|rrrrrrr|}
\hline
Parameter & $\mu$ & $M_{2}$ & $\tan \beta$ & $m_{A}$ & $m_{0}$ &
$A_{b}/m_{0}$ & $A_{t}/m_{0}$ \\
Unit & GeV & GeV & 1 & GeV & GeV & 1 & 1 \\ \hline Min & -50 000 &
-50 000 & 1.0 & 0        & 100 & -3 & -3 \\
Max & 50 000 & 50 000 & 60.0 & 10 000 & 30 000 & 3 & 3 \\ \hline
\end{tabular}
\end{center}
\label{tab:scans}
\end{table}

\begin{figure}[t]
\epsfxsize=16.4pc
\epsfbox{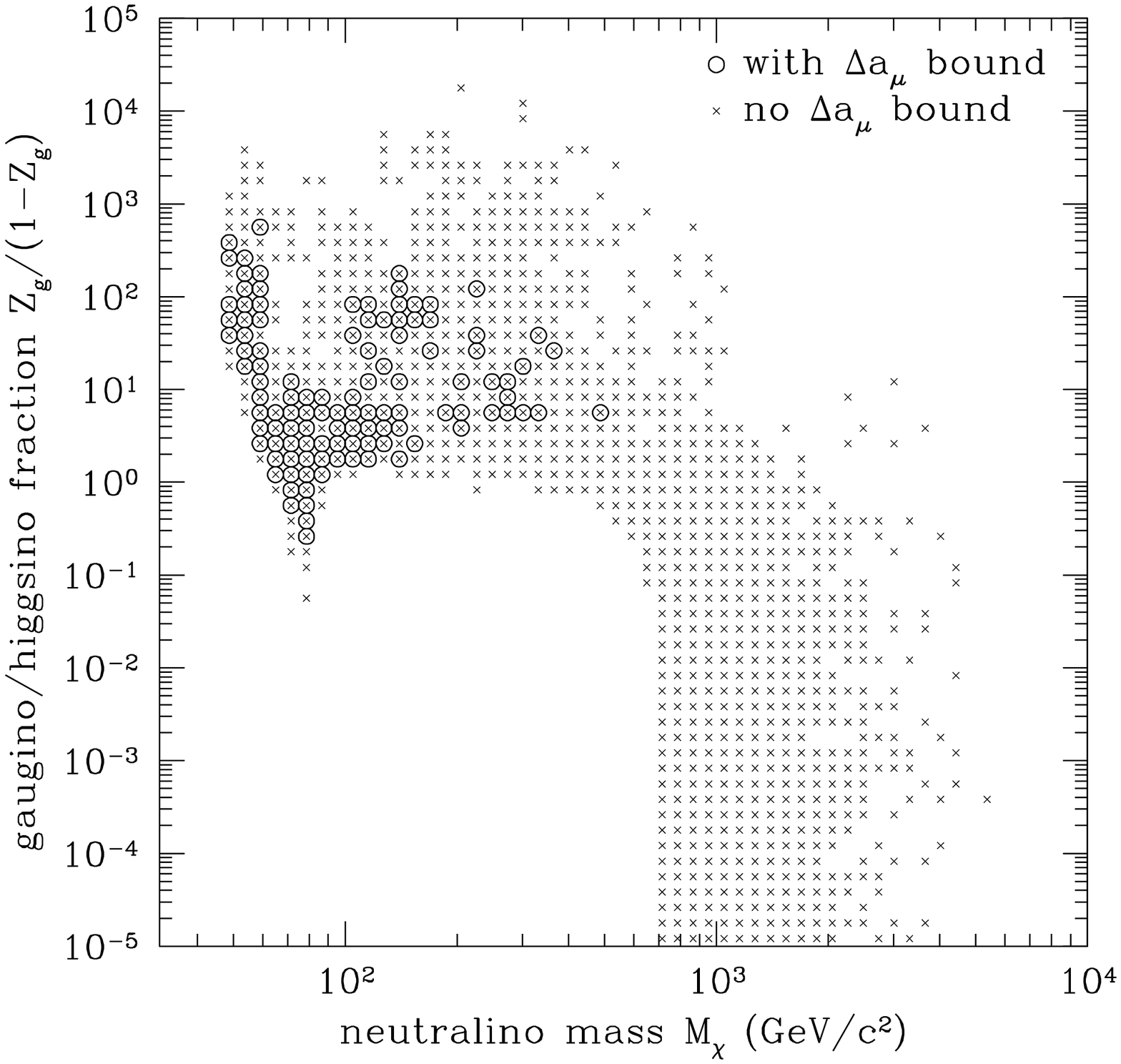}
\epsfxsize=16.4pc
\epsfbox{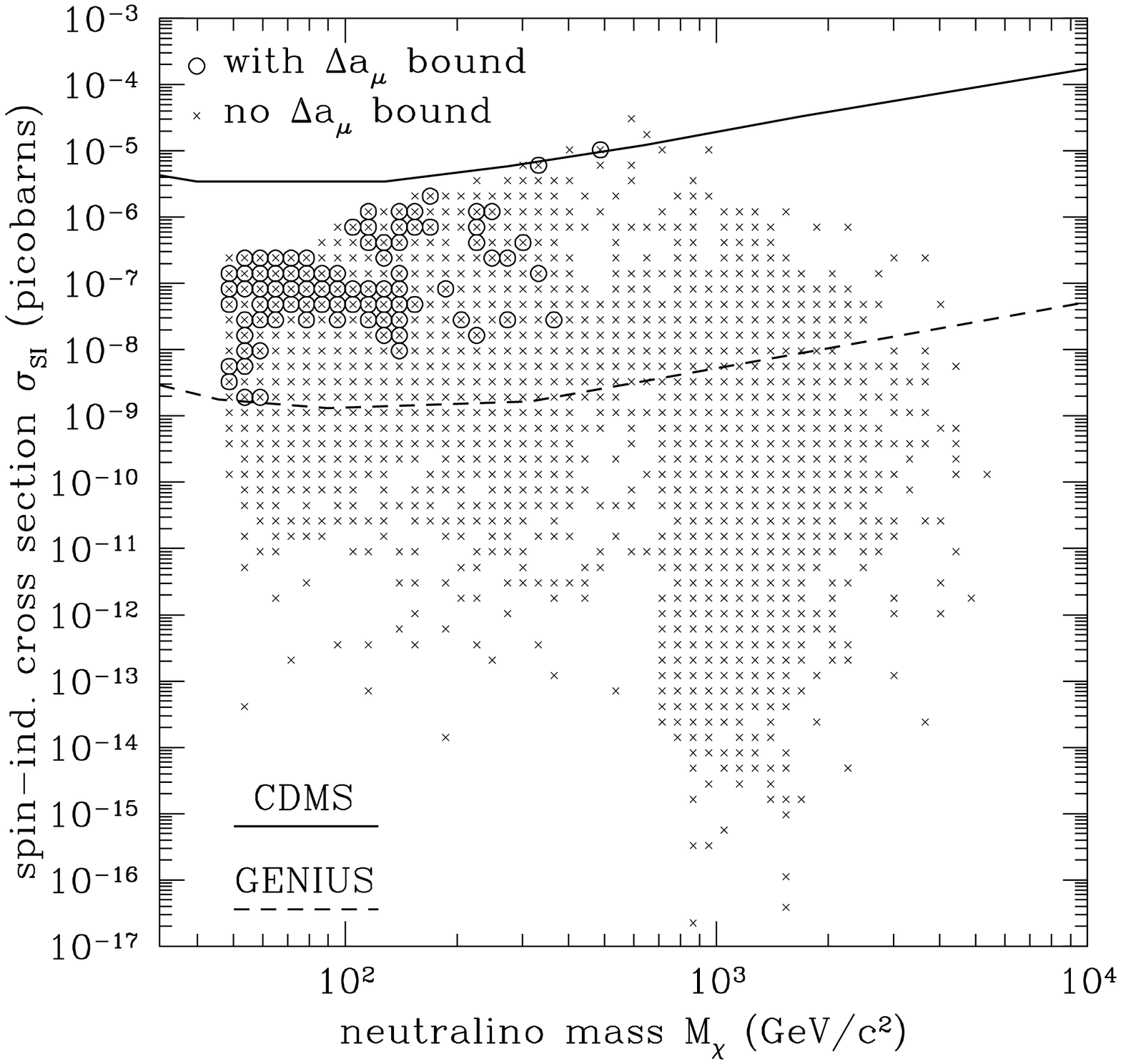}\\
\epsfxsize=16.4pc
\epsfbox{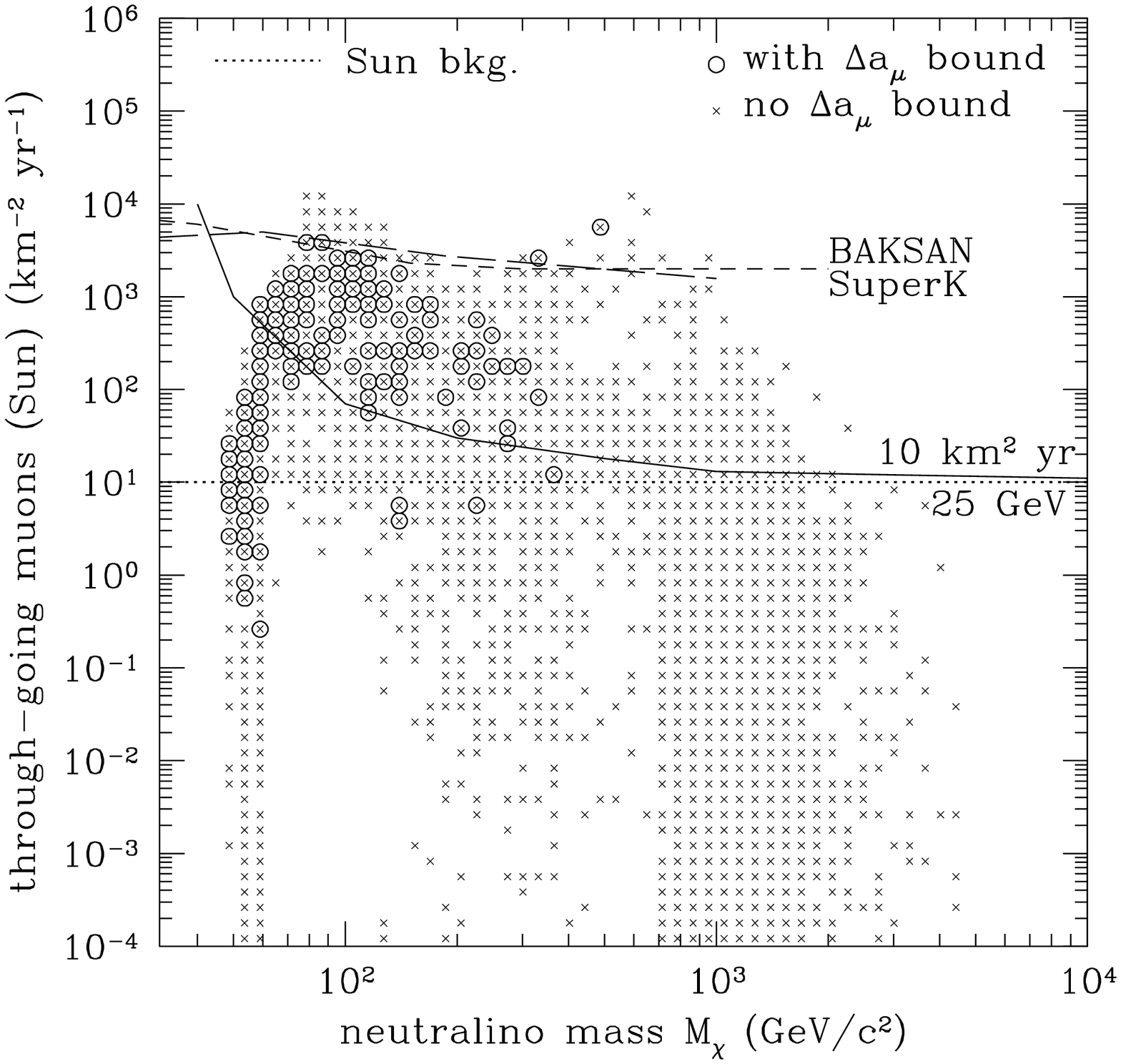}
\epsfxsize=16.4pc
\epsfbox{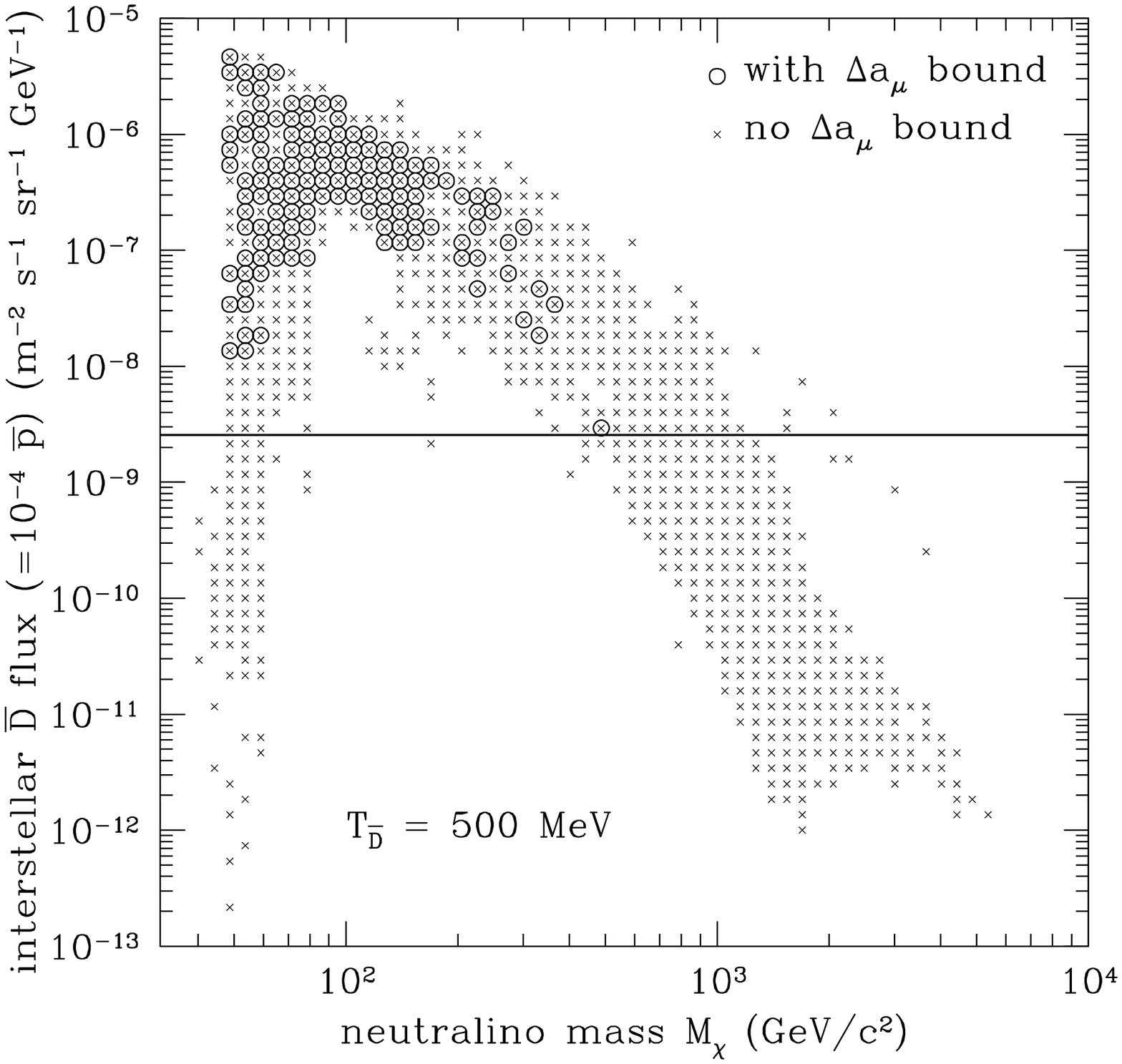}
\caption{Scan of supersymmetric parameter space, before and after the $\dams$
constraint.  Allowed models are plotted with circles.  In the top left, we plot
the gaugino / higgsino ratio against the neutralino mass.  In the top right, we
plot the elastic scattering cross section with present and future experimental
reach.  In the bottom left we plot the muon flux induced by neutrinos from the
Sun, again with present and future experimental reach.  In the bottom right we
plot an estimate of the antideuteron flux, with future experimental reach.}
\label{fig:detect}
\end{figure}

\section{Dark Matter Detection}

The most pronounced effect of applying the $\dams$ bound is an upper limit of
about 500 GeV on the neutralino mass. The previous bound of 7 TeV was
cosmological\cite{coann}, that is from the constraint $\Omega_\chi h^2<1$.
Furthermore, the neutralino must have at least a 10\% admixture of gauginos,
namely neutralino dark matter can not be very purely higgsino-like, as seen in
the top left panel of Fig.~\ref{fig:detect}.

Neutralinos in the galactic halo are constantly passing through the Earth, and
may be detectable with sensitive underground instruments such as
CDMS\cite{cdms} and DAMA\cite{dama}. The neutralino--nucleon elastic scattering
cross section is correlated with $\dams$\cite{drees}. In the top right panel of
Fig.~\ref{fig:detect}, we plot the spin-independent neutralino-proton
scattering cross section.  The constraint due to $\dams$ places a bound that is
conceivably detectable in future experiments, such as GENIUS\cite{genius}.

Another possible method to detect neutralino dark matter is neutrino
telescopes, such as at Lake Baikal\cite{baikal}, Super-Kamiokande\cite{superk},
in the Mediterranean\cite{antares}, and the south pole\cite{amanda}.
Neutralinos in the galactic halo undergo scatterings into bound orbits around
the Earth and Sun, and subsequently sink to the centers of these bodies and
annihilate, producing a neutrino signal at GeV and higher energies.  To
illustrate, we plot the rate of neutrino-induced through-going muons from the
Sun, along with the unsubtractable background, in the bottom left panel of
Fig.~\ref{fig:detect}.  We see that the $\dams$ bound removes most undetectable
models, except for some with threshold effects\cite{neutrate}.

One final interesting possibility is that neutralino annihilation in the
galactic halo can produce an observable flux of antideuterons\cite{dbar}.  At
low energies, the background should be smaller than for antiprotons, and a
signal may be detectable in future experiments\cite{dbarexp}, as seen in the
bottom right of Fig.~\ref{fig:detect}.

\end{document}